# Inverse Transformation Optics and Reflection Analysis for Two-Dimensional Finite Embedded Coordinate Transformation

Pu Zhang, Yi Jin, and Sailing He, *Senior Member, IEEE*

*Abstract*—Inverse transformation optics is introduced, and used to calculate the reflection at the boundary of a transformation medium under consideration. The transformation medium for a practical device is obtained from a two-dimensional (2D) finite embedded coordinate transformation (FECT) which is discontinuous at the boundary. For an electromagnetic excitation of particular polarization, many pairs of original medium (in a virtual space $V'$) and inverse transformation can give exactly the same anisotropic medium through the conventional procedure of transformation optics. Non-uniqueness of these pairs is then exploited for the analysis and calculation of the boundary reflection. The reflection at the boundary of the anisotropic FECT medium (associated with the corresponding vacuum virtual space $V$) is converted to the simple reflection between two isotropic media in virtual space $V'$ by a selected inverse transformation continuous at the boundary. A reflectionless condition for the boundary of the FECT medium is found as a special case. The theory is verified numerically with the finite element method.

*Index Terms*-- inverse transformation optics, finite embedded coordinate transformation, reflection, reflectionless condition.

## I. INTRODUCTION

Since the appearance of electromagnetic (EM) cloaking [1] based on transformation optics [2], optical invisibility [3-10] and agile control of EM waves [11-19] have become of great interest worldwide. Being one of the ultimate illusions in science fictions, invisibility promises many novel phenomena and special applications. In order to achieve this, a coordinate transformation opening a point or a line to a concealment space is used to design an EM cloaking shell with transformation optics. In the shell, EM waves are guided around the concealment volume, leaving the external field distribution unperturbed as if nothing exists inside. However, we need to fabricate these anisotropic and inhomogeneous artificial metamaterials [20]. Besides transformation optics, many other mechanisms have also been explored for invisibility: For a small dielectric particle, if coated with a plasmonic shell, the total scattering could be drastically reduced due to the cancellation of opposite local polarizabilities [6]; In a quasi-static limit, a polarizable dipole placed near a superlens can give anomalous resonance, leading to some invisibility effects [7]; With the approximation of geometrical optics, a conformal mapping technique was utilized to create optical cloaking [8,9]; More recently, complementary media and transformation optics were combined to realize optical transparency at a distance [10].

As a theoretical tool, transformation optics itself has also become a hot topic in the community of optics, microwave, and acoustics, etc. Due to the natural invariance of Maxwell's equations under a coordinate transformation, transformation optics allows people to control the geometry of light (through a particular coordinate transformation in a virtual space) with a physical material distribution (transformation medium). Apart from EM cloaking, transformation optics also enables various novel devices for EM wave manipulations, including some field concentrator [13], rotator [14], hyperlens [15], and arbitrary waveguide transition [16]. It is noted, however, coordinate transformations continuous at the boundaries only control EM waves within the devices under consideration. Later, transformations discontinuous at the boundaries (specification "*at the boundaries*" may be omitted hereafter for simplicity) were used to enable EM wave manipulations outside the devices, bringing new applications such as some beam shifter [12], expander [17], polarization rotator [18], and phase transformer [19], etc. Reflectionless nature at the boundaries of a device is usually preferred. Discontinuous transformations usually give some reflection at the boundaries of the FECT medium [12], and reflectionless conditions should therefore be investigated. This issue has been studied in two very recent works [21,22]. In Ref. [21], the reflectionless condition for a transformation medium was fulfilled by matching the boundary condition for the incident and transmitted fields (and consequently the reflected field will be zero). Generic boundary conditions between two generalized transformation media were discussed in Ref. [22], where some four-dimensional formulation is adopted to include the time variable in the transformation.

Manuscript received June 5, 2009. The partial support of the National Basic Research Program (973) of China (under Project No. 2004CB719800), the National Natural Science Foundation of China (Grant No. 60688401), and a Swedish Research Council (VR) grant on metamaterials is gratefully acknowledged.

All authors are with Centre for Optical and Electromagnetic Research, State Key Laboratory of Modern Optical Instrumentation, Zhejiang University, Zijingang, 310058 Hangzhou, PR China. P. Zhang and S. He (e-mail: sailing@kth.se) are also with Department of Electromagnetic Engineering, Royal Institute of Technology, S-10044 Stockholm, Sweden.

In this paper, inverse transformation optics is introduced first. For a specific polarization in two dimensions, a general anisotropic medium can be viewed as a transformation medium obtained from a pair of an original medium (in a virtual space *V'*) and an inverse transformation. Such a pair (associated to the same anisotropic medium) is not unique. When applied to some reflection analysis for an FECT medium (anisotropic in this case), we select among these pairs a specific pair with an inverse transformation continuously passing the boundary to calculate the reflection at the boundary. In this way the problem is simplified to the reflection between two isotropic media in virtual space *V'*. The reflection at the boundary of the FECT medium is then deduced directly from the reflection in virtual space *V*. A reflectionless condition is obtained as a special case. Numerical simulations based on the finite element method are also carried out for validation and visualization.

## II. INVERSE TRANSFORMATION OPTICS AND REFLECTION ANALYSIS

In principle, transformation optics is based on the form-invariance of Maxwell's equations under coordinate transformations. Generally speaking, the invariance holds in four-dimensional Minkowski space, but only spatial variables are considered in a practical design. Consider a virtual space with an EM field distribution $(\bar{E}, \bar{H})$ and an original medium characterized by $(\bar{\bar{\varepsilon}}, \bar{\bar{\mu}})$. A relation between the virtual space and the physical space (whose variables are primed) can be established through the following spatial coordinate transformation

$$(x', y', z') = [X(x,y,z), Y(x,y,z), Z(x,y,z)]. \quad (1)$$

Due to the invariance of Maxwell's equations, the following transformed quantities in the physical space also satisfy Maxwell's equations

$$\bar{\bar{\varepsilon}}' = \bar{\bar{J}}\bar{\bar{\varepsilon}}\bar{\bar{J}}^T / \det(\bar{\bar{J}}), \quad \bar{\bar{\mu}}' = \bar{\bar{J}}\bar{\bar{\mu}}\bar{\bar{J}}^T / \det(\bar{\bar{J}}), \quad (2)$$

$$\bar{E}' = (\bar{\bar{J}}^T)^{-1} \bar{E}, \quad \bar{H}' = (\bar{\bar{J}}^T)^{-1} \bar{H}, \quad (3)$$

where $\bar{\bar{J}} = \partial(x',y',z')/\partial(x,y,z)$ is the Jacobian matrix of the coordinate transformation. Thus, the transformed quantities can be given the physical meanings describing EM interactions with the transformation medium. In the design of a practical device, a specific transformation will be chosen to fulfill some desired functionalities.

Transformation optics allows the embodiment of prescribed complex coordinate transformations in the physical space, resulting in inhomogeneous and anisotropic transformation media. On the contrary, an EM problem in a virtual space is usually easy to solve. Then if a problem is formulated in the virtual space, we can solve the problem easily and then transform the solution in the virtual space back to the physical space. Below, we introduce the framework of inverse transformation optics for the 2D case (all physical quantities are assumed to be *z*-invariant). Original media in virtual spaces are supposed to be isotropic, so that problems become simple enough there. Since some birefringence phenomenon occurs in anisotropic media but not in isotropic media, it is difficult to transform generally an anisotropic medium back to an isotropic medium. In the 2D case, we can treat the two polarization states of EM waves separately. The TE polarization ($E_z$ polarized) case is formulated explicitly here, and the results for the TM polarization can be deduced in the same way. The general anisotropic medium in physical space *P* is described by *ε'* (zz-element of the permittivity tensor $\bar{\bar{\varepsilon}}'$; the other elements of $\bar{\bar{\varepsilon}}'$ have no effect to the TE polarization in 2D) and permeability tensor

$$\bar{\bar{\mu}}'_t = \begin{pmatrix} \mu'_{xx} & \mu'_{xy} \\ \mu'_{xy} & \mu'_{yy} \end{pmatrix}.$$

In the framework of inverse transformation optics, the anisotropic medium should be interpreted as a transformation medium. Correspondingly, we assume that an original medium in virtual space *V'* is characterized by *ε* and *μ*, and the coordinate transformation is described by

$$(x', y', z') = [X(x,y), Y(x,y), Z(z)]. \quad (4)$$

These assumptions are temporarily universal. Specific constraints will be imposed later in the paper. According to transformation optics, Eq. (2) should be satisfied under the above assumptions. Here we use the following variable replacements for brevity of mathematical derivation and expression,

$$\frac{\partial X}{\partial x} = A\cos\alpha, \frac{\partial Y}{\partial x} = B\cos\beta,$$
$$\frac{\partial X}{\partial y} = A\sin\alpha, \frac{\partial Y}{\partial y} = B\sin\beta. \quad (5)$$

With Eq. (5), the first four elements of the Jacobian matrix

$$\bar{\bar{J}} = \begin{pmatrix} \frac{\partial X}{\partial x} & \frac{\partial X}{\partial y} & 0 \\ \frac{\partial Y}{\partial x} & \frac{\partial Y}{\partial y} & 0 \\ 0 & 0 & \frac{\partial Z}{\partial z} \end{pmatrix} \quad (6)$$

for the transformation in Eq. (4) are then expressed in terms of four new variables *A*, *B*, *α* and *β*. After some algebraic manipulations, we obtain the following expressions from Eq. (2)

$$A = \sqrt{\frac{\varepsilon\mu}{\varepsilon'\mu'_{yy}}}\bigg/\sin(\beta-\alpha),$$

$$B = \sqrt{\frac{\varepsilon\mu}{\varepsilon'\mu'_{xx}}}\bigg/\sin(\beta-\alpha),$$

$$\frac{\partial Z}{\partial z} = \frac{\mu}{\sqrt{\mu'_{xx}\mu'_{yy}}}\bigg/\sin(\beta-\alpha), \qquad (7)$$

$$\cos(\beta-\alpha) = \frac{\mu'_{xy}}{\sqrt{\mu'_{xx}\mu'_{yy}}}.$$

One sees that the specific anisotropic medium can be viewed as a transformation medium associated with a pair of an original medium (in virtual space $V'$) and an inverse transformation. Such a pair is denoted by $[(\varepsilon,\mu),\bar{\bar{J}}^{-1}]$. The anisotropic medium can be generated with such a pair through the conventional procedure of transformation optics. In this sense, we define inverse transformation optics as a sort of mapping from the anisotropic medium to a pair $[(\varepsilon,\mu),\bar{\bar{J}}^{-1}]$ of an original medium and an inverse transformation. Inverse transformation optics is indispensable for the integrity of the whole transformation optics theory. Thus, we need to reverse the procedure for transformation optics (where a transformation is prescribed to obtain an anisotropic transformation medium). Using the present inverse transformation optics, problems involving anisotropic media can be simplified greatly. Solutions to complicated problems are then found by transforming the solutions to some simple problems. The above equation also implies that for a specific anisotropic transformation medium the pair of an original medium and an inverse transformation is not determined uniquely (due to the freedom of $\alpha$ and $\beta$). Many different possible pairs of original medium and inverse transformation can give the same specific anisotropic medium. The non-uniqueness of the solution pair to the problem of inverse transformation optics will be exploited below for the reflection analysis at the boundary of the FECT medium.

In a device design using transformation optics, reflectionless feature (or very small reflection) at the boundary of the device is usually preferred. Analysis of boundary reflection for such a device is therefore necessary. Fig. 1 illustrates two types of transformations between a virtual space and the physical space. Here we only consider at one boundary (denoted by a vertical dashed axis) of a device. The transformation medium (denoted by $M$) for a practical device is designed with a FECT (a specific discontinuous transformation denoted by $T$) and its corresponding virtual space is usually assumed to be vacuum. Thus reflection causing some adverse effects may arise due to the discontinuity of the FECT $T$. Calculation of the reflection level is thus necessary in order to evaluate the device performance. To avoid dealing directly with the complex transformation medium, the approach of transformation is helpful. However, in order to establish a mapping between virtual space $V$ and physical space $P$ in the whole domain, continuity condition of EM fields at the boundary requires the FECT $T$ be identity transformation (and thus continuous) at the boundary (cf. Eq. (3)). Thus discontinuous FECT $T$ itself can not establish a mapping in the whole domain (but is only valid in a finite domain, thus called finite embedded coordinate transformation), and consequently is not convenient for the reflection analysis. Using the above inverse transformation optics, the general anisotropic medium (now fixed as the FECT medium $M$) can be generated from many possible pairs of original medium in virtual space $V'$ and inverse transformation. However, these inverse transformations obtained from the inverse transformation optics procedure are not guaranteed to be continuous, and those discontinuous inverse transformations can not be used directly for the reflection analysis. Among these pairs, we try to select a special pair with its inverse transformation continuous at the boundary to establish a useful mapping between virtual space $V'$ and physical space $P$. In virtual space $V'$, the reflection at the interface of two isotropic media can be easily calculated.

Note that one should not be confused with the two different types of mapping transformations (continuous and discontinuous) used in the present paper. One mapping type is the Finite Embedded Coordinate Transformation (FECT), which is a specific discontinuous transformation (from virtual space $V$ to physical space $P$) giving the transformation medium for a practical device. The other mapping type is of inverse transformations (from physical space $P$ to virtual space $V'$), which are obtained from inverse transformation optics procedure, including both continuous and discontinuous ones (but only the continuous one is selected for our reflection analysis). The two different types of transformations and their properties are distinguished and emphasized under different names in the present paper: "FECT" (discontinuous for a practical device) and "inverse transformation" (continuous in our reflection analysis).

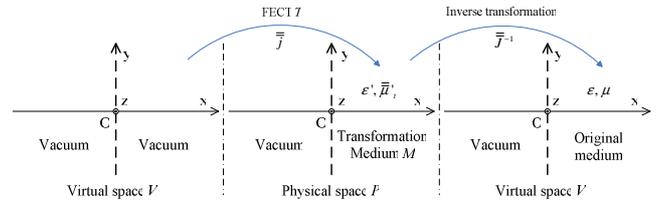

Fig. 1. Two types of transformations (for $E_z$-polarization in 2D) between a virtual space and the physical space near a boundary of a device designed with the transformation optics.

Practical devices designed by transformation optics are usually inhomogeneous, even at the boundaries. Global treatment of the reflection is therefore complicated and usually unnecessary for transformation optics applications. Instead, a local view is shown to be convenient and sufficient to identify the reflection at any point on the boundaries. Thus even inhomogeneity exists along the boundaries the reflection can be evaluated point by point. For example, the medium at points on the boundaries may be parameterized so that the complete information of the reflection at the boundaries can be described with the help of the parameter. As for points inside an FECT medium, since virtual space $V$ associated with the FECT is vacuum and the FECT is continuous except at the boundaries, no reflection will be induced at any point inside the FECT medium. Under the assumption of the local view, coordinate transformations can be expanded into Taylor series at a boundary, and terms of orders higher than one can be omitted. Hence, the FECT and inverse coordinate transformations in Eq. (4) become linear at a boundary. Correspondingly, the FECT medium $M$ near the boundary can be treated locally as a homogeneous one. Thus, we can make the reflection analysis for the case of a (locally) homogeneous transformation medium. Considering the configuration shown in Fig. 1, the FECT medium $M$ is then transformed back to an original medium in virtual space $V'$. Since a continuous transformation can give a mapping between virtual space $V'$ and physical space $P$, we consider a particular inverse transformation continuous at the boundary ($X(0,y) = 0$, $Y(0,y) = y$, $Z(z) = z$), i.e.,

$$\frac{\partial X}{\partial y} = 0, \frac{\partial Y}{\partial y} = \frac{\partial Z}{\partial z} = 1. \quad (8)$$

Inserting Eqs. (5) and (7) into Eq. (8), the continuous inverse transformation and the original isotropic medium are determined

$$\sin\alpha = 0, \cos\beta = \frac{\mu'_{xy}}{\sqrt{\mu'_{xx}\mu'_{yy}}}, \varepsilon = \sqrt{\frac{\mu'_{xx}}{\mu'_{yy}}}\frac{\varepsilon'}{\sin\beta}, \quad (9)$$

$$\mu = \sqrt{\mu'_{xx}\mu'_{yy}}\sin\beta.$$

Note that we still adopt $\varepsilon'$ and $\bar{\bar{\mu}}'_t$ to characterize the anisotropic FECT medium $M$. They can be expressed with the Jacobian matrix of the FECT $T$ when needed. For the fixed FECT medium $M$, the original medium in virtual space $V'$ under the continuous inverse transformation must be fixed with the material parameters given by Eq. (9). Jacobian matrix $\bar{\bar{J}}$ associated to the continuous inverse transformation is obtained by substituting the above results into Eq. (6). The reflection at the boundary of the FECT medium $M$ can be calculated from the reflection between the vacuum and the isotropic original medium in virtual space $V'$. Since both the reflected and incident waves are on the vacuum side, where the transformation is an identity transformation, the reflected and incident waves in physical space $P$ are the same as their counterparts in virtual space $V'$. Thus, the reflection coefficient, which is the ratio between the two waves, in physical space $P$ is identical to that in virtual space $V'$. Specifically, if the incident wave is described as $E'_{Inc} = e^{ik_0(x'\cos\theta + y'\sin\theta) - i\omega t}$, the reflected wave can be described as

$$E'_R = R e^{ik_0(-x'\cos\theta + y'\sin\theta) - i\omega t}, \quad (10)$$

where $k_0$ is the angular wavenumber in vacuum. Reflection coefficient $R$ is readily computed in virtual space $V'$ with the following Fresnel formula

$$R = (1-p)/(1+p), \quad (11)$$

where

$$p = \sqrt{\frac{\varepsilon}{\mu}}\frac{\cos\gamma}{\cos\theta},$$

and the refraction angle is determined by

$$\sin\gamma = \sin\theta / \sqrt{\varepsilon\mu}.$$

Although transmission analysis is not the topic of the present paper, transmission can also be treated similarly with the present inverse transformation optics. In virtual space $V'$, the transmitted field is easy to obtain. The transmitted field in physical space $P$ can then be obtained from that in virtual space $V'$ (on the original medium side) through Jacobian matrix $\bar{\bar{J}}$ associated to the continuous inverse transformation. In this way, the transmission coefficient is calculated easily. If the Jacobian matrix of the FECT $T$ is fixed to $\bar{\bar{j}}$, the material parameters of the FECT medium $M$ are calculated through Eq. (2) from vacuum virtual space $V$

$$\varepsilon' = \frac{j_{zz}^2}{\det(\bar{\bar{j}})}, \mu'_{xx} = \frac{j_{xx}^2 + j_{xy}^2}{\det(\bar{\bar{j}})},$$

$$\mu'_{xy} = \frac{j_{xx}j_{yx} + j_{xy}j_{yy}}{\det(\bar{\bar{j}})}, \mu'_{yy} = \frac{j_{yx}^2 + j_{yy}^2}{\det(\bar{\bar{j}})}. \quad (12)$$

Substituting Eq. (12) into Eq. (9), we obtain

$$\sin\beta = \frac{j_{xx}j_{yy} - j_{xy}j_{yx}}{\sqrt{(j_{xx}^2 + j_{xy}^2)(j_{yx}^2 + j_{yy}^2)}},$$

and the material parameters of the isotropic original medium can then be expressed in terms of the Jacobian matrix elements as follows

$$\varepsilon = \frac{j_{zz}(j_{xx}^2 + j_{xy}^2)}{(j_{xx}j_{yy} - j_{xy}j_{yx})^2}, \mu = \frac{1}{j_{zz}}. \quad (13)$$

Reflection coefficient $R$ is then calculated from Eq. (11) with

$$p = \frac{j_{zz}\sqrt{j_{xx}^2 + j_{xy}^2}}{j_{xx}j_{yy} - j_{xy}j_{yx}}\frac{\cos\gamma}{\cos\theta}. \quad (14)$$

According to the above reflection analysis using the inverse transformation optics, we can easily obtain a reflectionless condition in terms of the FECT $T$ itself simply by requiring a vanishing reflection coefficient for an arbitrary incidence angle. In virtual space $V'$, zero reflection for any incidence angle occurs only when

$$\varepsilon = \mu = 1. \quad (15)$$

Substituting Eq. (13) into Eq. (15), reflectionless condition Eq. (15) becomes

$$\frac{j_{xx}^2 + j_{xy}^2}{(j_{xx}j_{yy} - j_{xy}j_{yx})^2} = j_{zz} = 1. \quad (16)$$

All our studies are associated to the schematic configuration shown in Fig. 1. Obviously coordinates $x$ and $y$ are not equivalent in the above results. In Ref. [21] the reflectionless condition was fulfilled by matching the boundary condition for the incident and transmitted fields, and no explicit expression (like Eq. (16) in our paper) for the zero reflection was given.

Consequently their theory can not confirm directly zero reflection even when one knows the Jacobian matrix of transformation *T* for a specific device. In the present approach of inverse transformation optics, our reflectionless condition (Eq. (16)) is explicit and simple (and yet sufficient and necessary), and allows one to determine quickly (from the Jacobian matrix of the FECT *T*) whether the specific FECT medium *M* is reflectionless at the boundary (and thus is much more convenient for use as compared with the one given in Ref. [21]). If it is not satisfied, the non-zero reflection can also be evaluated quantitatively. Furthermore, the present inverse transformation optics provides an alternative understanding of the reflectionless phenomenon at the boundary of the designed FECT medium. For this FECT medium *M*, inverse transformation optics can give many pairs of isotropic original medium (in virtual space *V'*) and inverse transformation. Since only an inverse transformation continuous at the boundary can give a mapping (from physical space *P* to virtual space *V'*) in the whole domain (as explained earlier), we select a particular inverse transformation continuous at the boundary for the reflection analysis. Then, a unique pair (cf. Eq. (9)) is selected among all these pairs. If the reflectionless condition is satisfied, the isotropic original medium is vacuum (cf. Eq. (13)). Then obviously no reflection will occur. Otherwise, the isotropic original medium different from vacuum will cause some reflection.

### III. NUMERICAL SIMULATION AND DISCUSSIONS

In this section, numerical simulations are carried out with finite-element-method based commercial package COMSOL [23] to validate our theory and conclusions. As a numerical example, we choose the following Jacobian matrix for the FECT *T*

$$\bar{\bar{j}} = \begin{pmatrix} 1 & \frac{3}{4}+\delta & 0 \\ 1 & 2 & 0 \\ 0 & 0 & 1 \end{pmatrix}.$$

Material parameters of this FECT medium *M* is calculated from Eq. (12)

$$\varepsilon' = 1/\left(\tfrac{5}{4}-\delta\right),$$

$$\bar{\bar{\mu}}'_t = \begin{pmatrix} 1+\left(\tfrac{3}{4}+\delta\right)^2 & \tfrac{5}{2}+2\delta \\ \tfrac{5}{2}+2\delta & 5 \end{pmatrix}/\left(\tfrac{5}{4}-\delta\right).$$

When $\delta = 0$, reflectionless condition (16) is satisfied. In our simulations, the computational domain surrounded by perfectly matched layers is separated into two parts (vacuum and the FECT medium). A 2D Gaussian beam of TE polarization impinging at an incidence angle of 45° is used as the excitation source to illustrate the reflectionless feature when $\delta$ = 0. Resultant distribution of the electric field is shown in Fig. 2. Arrow plot of time averaged power flow is also included in Fig. 2, indicating total transmission without any reflection.

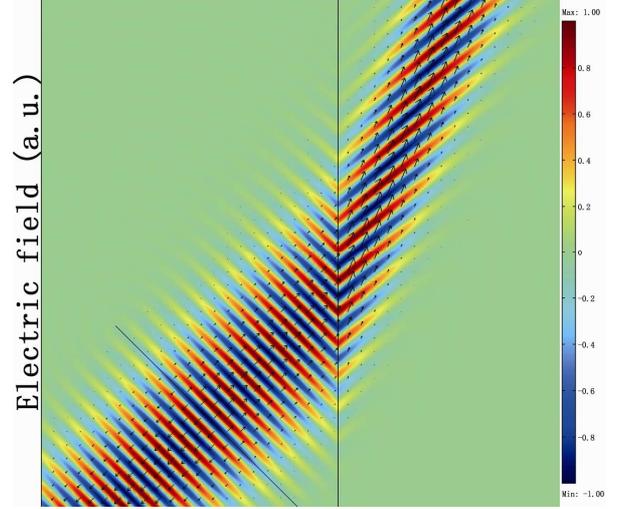

Fig. 2. The distribution of the electric field for the interaction at the boundary of the transformation medium (with $\delta = 0$) for a TE polarized Gaussian beam.

When $\delta \neq 0$ (i.e., the reflectionless condition is not satisfied), nonzero reflection occurs. The reflection can be predicted by Eqs. (11) and (14). The settings of the computational domain are the same as those in the previous simulation. Let $\delta$ vary in the range of [-10, 5/4)**.** The incidence angle is kept 45° as in the previous simulation. Parameters of the original medium in virtual space *V'* of the selected pair can help to understand well the reflection. Using Eq. (13), $\mu = 1$ is found invariant with $\delta$, while

$$\varepsilon = \frac{j_{zz}\left(j_{xx}^2 + j_{xy}^2\right)}{\left(j_{xx}j_{yy} - j_{xy}j_{yx}\right)^2}$$

is evaluated numerically. Both the reflection coefficient (norm) and the permittivity of the original medium are shown in Fig. 3 as deviation $\delta$ varies. When the deviation is zero, we see the permittivity is 1 and the reflection is zero, in agreement with the previous simulation. In the middle region total internal reflection occurs due to the low permittivity (also shown with the dashed line in Fig. 3) of the original medium. In the limit of $\delta \rightarrow -\infty$, the original medium tends to vacuum with the reflection decreasing to zero.

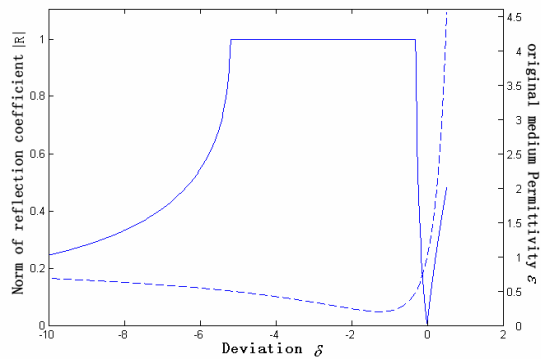

Fig. 3. The permittivity of the original medium (dashed line) and the norm of the reflection coefficient (solid line) at the boundary of the FECT medium (with $\delta \in$ [-10, 5/4)) for a TE polarized Gaussian beam with an incidence angle of 45º.

It is seen that the reflection coefficient increases rather quickly with a small deviation $\delta$. As an example, we choose a small deviation of $\delta = -0.25$ and obtain a nonzero reflection coefficient of 0.5. The distribution of the electric field norm is shown in Fig. 4 for verification. Furthermore, the Jacobian matrix associated to the continuous inverse transformation becomes

$$\bar{\bar{J}} = \begin{pmatrix} A & 0 & 0 \\ B\cos\beta & 1 & 0 \\ 0 & 0 & 1 \end{pmatrix},$$

where $A$, $B$ and $\cos\beta$ can be calculated from Eqs. (7) and (9).

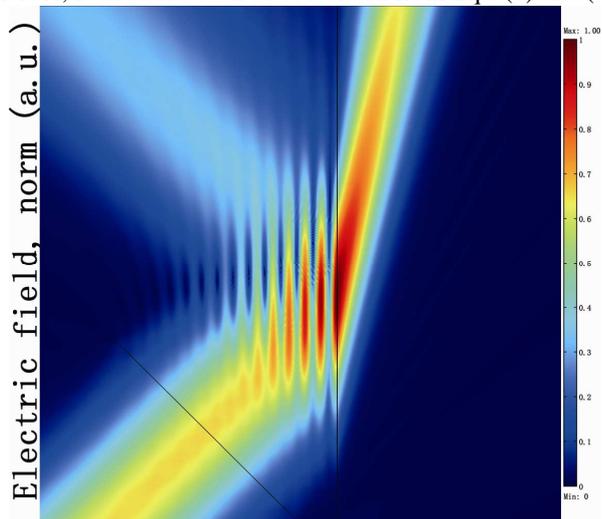

Fig. 4. The distribution of the norm of the electric field for the interaction at the boundary of the FECT medium with a deviation of $\delta = -0.25$ when illuminated at an incidence angle of 45º by a TE polarized Gaussian beam.

## IV. CONCLUSIONS

In the present paper, we have defined the framework of inverse transformation optics and used it to analyze the reflection at a boundary of an FECT medium. For a specific polarization state in 2D, a general anisotropic medium can be viewed as a transformation medium. Many pairs of original medium (in virtual space *V'*) and inverse transformation can generate the anisotropic medium through the standard transformation optics. Among these pairs there exists a unique pair (associated with a continuous inverse transformation) that allows converting the reflection at the boundary of the FECT medium to that between two isotropic media in virtual space *V'*. Straightforward reflectionless condition for these two isotropic media is then transformed to physical space *P*, giving the reflectionless criterion for the FECT. Situations with non-zero reflection have also been treated using the inverse transformation optics. Numerical simulations based on the finite element method have been carried out to verify the proposed theory and conclusions. Although the regime is 2D here, the present method can also be effectively generalized in reflection analysis for some 3D FECT devices. The main contribution of the present paper is the introduction of inverse transformation optics, while the reflectionless condition is just a simple application example.

**Pu Zhang** received the B.Sc. degree from the Department of Optical Engineering of Zhejiang University, China, in 2005. Currently, he is working toward the Ph.D. degree, and his research activities are in the theory, design, and simulation of metamaterials and transformation optics.

**Yi Jin** received the M.Sc. and PhD degrees both from Zhejiang University, China, in 2003 and 2006, respectively. He was a post-doc at the Royal Institute of Technology (Sweden) during 2006 and 2007. Currently, he is a post-doc at the Department of Optical Engineering of Zhejiang University, China. His research interests include metamaterial, plasmonics and photonic crystals.

**Sailing He** (M'92–SM'98) received the Licentiate of Technology and the Ph.D. degree in electromagnetic theory from the Royal Institute of Technology, Stockholm, Sweden, in 1991 and 1992, respectively. Since then he has worked at the same division of the Royal Institute of Technology as an assistant professor, an associate professor, and a full professor. In 1999 Prof. He was also appointed by the Ministry of Education of China as a "Chang-jiang program" professor at Zhejiang University (China). He has first-authored one monograph (Oxford University Press) and authored/co-authored over 300 papers in refereed international journals (including about 100 papers in IEEE journals). Prof. He has given many invited/plenary talks in international conferences, and has served in the leadership for many international conferences. Prof. He is a Fellow of OSA (Optical Society of America) and SPIE, and a Topical Editor for Optics Letters.